\documentclass[12pt]{iopart}
\usepackage{bm,graphicx,amssymb}
\newcommand{\siesta}{\textsc{Siesta}}
\begin{document}

\title[Electronic properties of graphene antidot lattices]{Electronic properties of graphene antidot lattices}

\author{J A F\"{u}rst,$^1$ J G Pedersen,$^2$ C Flindt,$^3$ N A Mortensen,$^2$ M Brandbyge,$^1$ T G Pedersen$^4$ and A-P Jauho$^{1,5}$}
\address{$^1$ Department of Micro and Nanotechnology, Technical University of Denmark, DTU Nanotech, DTU-building 345 east, DK-2800 Kongens Lyngby, Denmark}
\address{$^2$ Department of Photonics Engineering, Technical University of Denmark, DTU Fotonik, DTU-building 343, DK-2800 Kongens Lyngby, Denmark}
\address{$^3$ Department of Physics, Harvard University, 17 Oxford Street, Cambridge, 02138 Massachusetts, USA}
\address{$^4$ Department of Physics and Nanotechnology, Aalborg University, DK-9220 Aalborg {\O}, Denmark}
\address{$^5$ Department of Applied Physics, Helsinki University of Technology, P.\ O.\ Box 1100, FI-02015 TKK, Finland}
\ead{joachim.fuerst@nanotech.dtu.dk}

\begin{abstract}
Graphene antidot lattices constitute a novel class of nano-engineered graphene devices with controllable electronic and optical properties. An antidot lattice consists of a periodic array of holes which causes a band gap to open up around the Fermi level, turning graphene from a semimetal into a semiconductor. We calculate the electronic band structure of graphene antidot lattices using three numerical approaches with different levels of computational complexity, efficiency, and accuracy. Fast finite-element solutions of the Dirac equation capture qualitative features of the band structure, while full tight-binding calculations and density functional theory are necessary for more reliable predictions of the band structure. We compare the three computational approaches and investigate the role of hydrogen passivation within our density functional theory scheme.
\end{abstract}



\maketitle

\section{Introduction}

Since its discovery in 2004 \cite{Novoselov2004,Novoselov2005}, graphene has become a research field of tremendous interest within the solid state physics community \cite{Neto2009}. The interest stems from the particular electronic properties of graphene as well as the promising perspectives for future technological applications \cite{Geim2009}. The electronic excitations around the Fermi level of graphene resemble those of massless, relativistic Dirac fermions, allowing predictions from quantum electrodynamics to be tested in a solid state system \cite{Katsnelson2006}. From a technological point of view, several future applications have already been envisioned. These include the use of graphene for single molecule gas detection \cite{Schedin2007}, graphene-based field-effect transistors \cite{Novoselov2004}, and quantum information processing in nano-engineered graphene sheets \cite{Trauzettel2007}. Additionally, graphene is the strongest material ever tested, suggesting the use of carbon-fiber reinforcements in novel material composites \cite{Lee2008}.

Metamaterials constitute another popular field of research in contemporary science. Contrary to conventional, naturally occurring materials, metamaterials derive their properties from their artificial, man-made, periodic small-scale structure rather than their chemical or atomic composition \cite{Pendry1996}. When properly designed and fabricated, metamaterials offer optimized and unusual, sometimes even counter-intuitive, responses to specific excitations \cite{Pendry2006}. Examples include metamaterials with negative permittivity and permeability \cite{Veselago1968}, superlenses \cite{Pendry2000,Fang2005}, and cloaking devices \cite{Schurig2006}. Photonic \cite{Yablonovitch1987,John1987} and phononic \cite{Vasseue2001} crystals are closely related to metamaterials, although they are typically designed to alter the response to electromagnetic and acoustic excitations, respectively, at wavelengths similar to the dimensions of the small-scale structure. The realization of artificial band structures in two-dimensional electron gasses may be pursued with similar approaches \cite{Flindt2005,Pedersen2008c}, allowing the formation of e.g.\ Dirac cones in conventional antidot lattices \cite{Park2009,Gibertini2009}.

Based on the above ideas, some of us have recently proposed to alter in a controllable manner the electronic and optical properties of graphene by fabricating a periodic arrangement of perforations or holes in a graphene sheet \cite{Pedersen2008}. We refer to this kind of structure as a graphene antidot lattice owing to its close resemblance with conventional antidot lattices defined on top of a two-dimensional electron gas in a semiconductor heterostructure \cite{Roukes1989,Ensslin1990}. Using tight-binding calculations we have shown that such a periodic array of holes in a graphene sheet causes a band gap to open up around the Fermi level \cite{Pedersen2008}, changing graphene from a semimetal to a semiconductor with corresponding clear signatures in the optical excitation spectrum \cite{Pedersen2008b}. Soon after our proposal, graphene antidot lattices were realized experimentally by Shen and co-workers \cite{Shen2008} and Eroms and Weiss \cite{Eroms2009} with lattice constants below 100 nm. The rapidly improving ability to pattern monolayer films with $e$-beam lithography suggests that graphene antidot lattices with typical dimensions towards the 10 nm scale may be within reach \cite{Han2007,Fischbein2008}. Furthermore, Girit and co-workers recently monitored the dynamics at the edges of a growing hole in real time using a transmission electron microscope \cite{Girit2009}, and Jia and co-workers demonstrated a method for producing graphitic nanoribbon edges in a controlled manner via Joule heating \cite{Jia2009}. Very recently, Rodriguez-Manzo and Banhart created individual vacancies in carbon nanotubes using a 1 \AA\ diameter $e$-beam \cite{Rodriguez-Manzo2009}. These advances suggest that fabrication of nano-scale graphene antidot lattices with desired hole geometries may be possible in the near future.

In the endeavors of modeling these structures one is faced with a compromise between computational efficiency and accuracy. Small-scale lattices with perfect periodicity and identical few-nm sized holes can be treated accurately with density functional theory (DFT), but this is a computationally heavy and time consuming approach, which limits the possibilities to perform large, systematic studies. For example, in order to model lattice disorder, such as variations in the hole geometry and alignment, it may be necessary to form a super cell containing several holes at the cost of an increased computational time. In order to circumvent this problem, one can make use of the pseudo-relativistic behavior of electrons in bulk graphene close to the Fermi level and solve the corresponding Dirac equation using computationally cheaper methods, however, possibly at the cost of a decreased computational accuracy.

The aim of this paper is to study the band structure of graphene antidot lattices using three numerical approaches of different computational complexity, efficiency, and accuracy. We first develop a computationally cheap scheme based on a finite-element solution of the Dirac equation. This method gives reasonable predictions for the size of the band gap due to the antidot lattice, but has limited accuracy in predicting the full band structure. For better predictions of the band structure, we employ a $\pi$-orbital tight-binding scheme, which is still numerically cheap and capable of treating larger antidot lattices. The results are compared with computationally demanding, full-fledged \emph{ab initio} calculations, based on density functional theory, which we expect to predict the band structure with the highest accuracy. The tight-binding calculations agree well with qualitative features of the band structure calculations based on density functional theory, although some differences are found on a quantitative level. Finally, we discuss hydrogen passivation along the edges of the holes in a graphene antidot lattice and study the influence on the electronic properties using density functional theory.

The paper is organized as follows: In Section \ref{sec:antidotlattice} we introduce graphene antidot lattices and give a brief overview of the existing literature on the topic. In Section \ref{sec:compmeth} we describe our three computational approaches; finite-element solutions of the Dirac equation (DE), a $\pi$-orbital tight-binding scheme (TB), and density functional theory calculations (DFT). A comparison and discussion of the results obtained using the three methods are given in Section \ref{sec:bandstruc}. Finally, we discuss in Section \ref{sec:passivation} the influence of hydrogen passivation on the band structure, before stating our conclusions in Section~\ref{sec:conclusions}.

\section{Graphene antidot lattices}
\label{sec:antidotlattice}

\begin{figure}
\begin{center}
\includegraphics[width=0.95\textwidth]{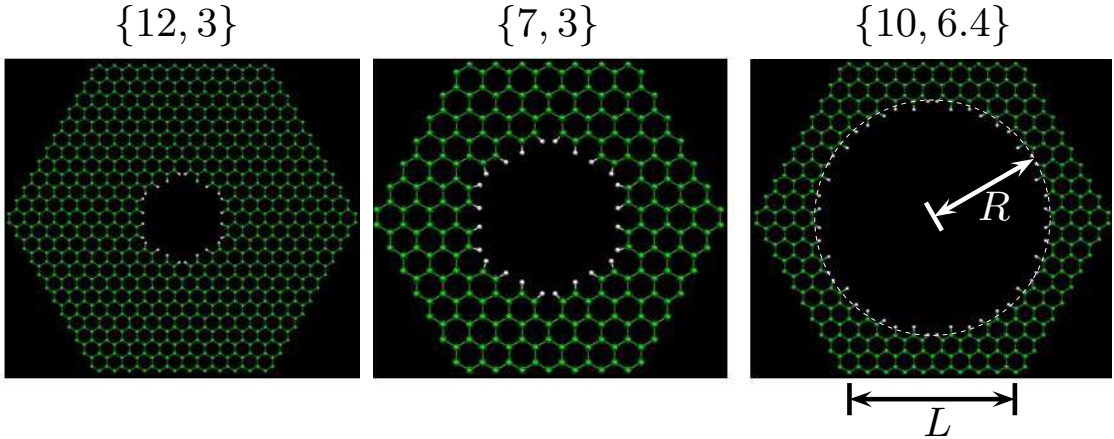}
\end{center}
\caption{Unit cells of three hexagonal graphene antidot lattices with different side lengths $L$ and hole radii $R$. The structures are denoted as $\{L,R\}$ with both lengths measured in units of the graphene lattice constant $a\simeq 2.46$ \AA. Here we have assumed that the edges of the holes have been hydrogen passivated (hydrogen shown as white atoms).}
\label{fig:unitcells}
\end{figure}

A graphene antidot lattice consists of a periodic arrangement of holes in a graphene sheet \cite{Pedersen2008}. In the following, we consider a hexagonal lattice of circular holes, but other lattice structures, e.g.\ square lattices, with different holes shapes are expected to exhibit similar physics. In particular, we anticipate an opening of a band gap around the Fermi level for a large class of antidot lattices \cite{Fuerst2009}. The hexagonal unit cells with different hole sizes are shown in Fig.\ \ref{fig:unitcells}. The structures are characterized by the side lengths $L$ of the hexagonal unit cells and the approximate radii $R$ of the holes, both measured in units of the graphene lattice constant $a=\sqrt{3}l_{C}\simeq 2.46$~\AA, where $l_{C}=1.42$~\AA\ is the bond length between neighboring carbon atoms. In Fig.\ \ref{fig:unitcells}, the holes are assumed to be passivated with hydrogen, using the bond length 1.1 \AA\ between neighboring carbon and hydrogen atoms. Throughout the paper, we denote a given structure as $\{L,R\}$, where $L$ is an integer, but $R$ not necessarily. We will consider only very small structures with $L\sim 10$. Although it may not be conceivable to fabricate such small structures within the near future, the small unit cells allow for systematic comparisons of our three computational schemes. In particular, with small unit cells we can perform computationally heavy DFT calculations. Importantly, simple scaling relations have been demonstrated for the size of the band gap in terms of the total number of atoms and the number of removed atoms within a unit cell, making it possible to extrapolate results to larger geometries \cite{Pedersen2008}. Such scaling relations may be helpful when modeling on-going experiments on graphene antidot lattices \cite{Shen2008,Eroms2009}.

In our original proposal for graphene antidot lattices, we focused on the possibility of fabricating intentional `defects' by leaving out one or more holes in the otherwise periodic structure \cite{Pedersen2008}. As we showed, such defects lead to the formation of localized electronic states at the locations of the defects with energies inside the band gap. Several such (possibly coupled) defects would then form a platform for coupled electronic spin qubits in a graphene-based quantum computing architecture \cite{Pedersen2008}. Similar ideas based on conventional antidot lattices defined on a two-dimensional gas in a semiconductor heterostructure have previously been studied by some of us \cite{Flindt2005,Pedersen2008c}. However, as already mentioned, the perfectly periodic graphene antidot lattice constitutes an interesting structure on its own. In particular, the controllable opening of a band gap may potentially pave the way for graphene-based semiconductor devices. In Ref.\ \cite{Pedersen2008b} some of us studied the optical properties of graphene antidot lattices, showing that they behave as dipole-allowed direct gap two-dimensional semiconductors with a pronounced optical absorption edge. Additional studies of the electronic properties have been performed by Vanevi\'{c}, Stojanovi\'{c}, and Kindermann \cite{Vanevic2009} as well as by some of us \cite{Fuerst2009}. Vanevi\'{c} and co-workers studied in detail the occurrence of flat bands due to sublattice imbalances and irregularities in the hole shapes at the atomic level. In our study, we addressed the roles of geometry relaxation and electron spin using DFT calculations. Very recently, Rosales and co-workers studied the transport properties of antidot lattices along graphene nanoribbons \cite{Rosales2009}. Turning around the ideas of making graphene semiconducting using periodic superlattices, it has recently been shown that periodic potential modulations may create graphene-like electronic band structures of two-dimensional gases in semiconductor heterostructures \cite{Park2009,Gibertini2009}. In that case, the possibility to control the slope of the linear bands and thus the velocity of the Dirac fermions is of great interest.

\section{Computational methods}
\label{sec:compmeth}

In the following we outline the three computational methods employed in this work. As a computationally cheap approach we consider first finite-element solutions of the Dirac equation (DE). Within this approach, large unit cells can be treated and the computations are fast. The method relies on the linear bands of bulk graphene around the Fermi level. As a more refined approach, we consider next $\pi$-orbital tight-binding calculations (TB). This method goes beyond the assumption of a linear band structure of bulk graphene, and the edges of the antidot holes can be carefully treated, including possible effects due to valley mixing. Finally, we consider full-fledged \emph{ab initio} calculations using DFT. While this method is computationally heavy, DFT is a widely used standard for doing first principles calculations and we expect it to provide the most detailed description of the electronic structure.

\subsection{Dirac equation (DE)}

\begin{figure}
\begin{center}
\includegraphics[width=\textwidth]{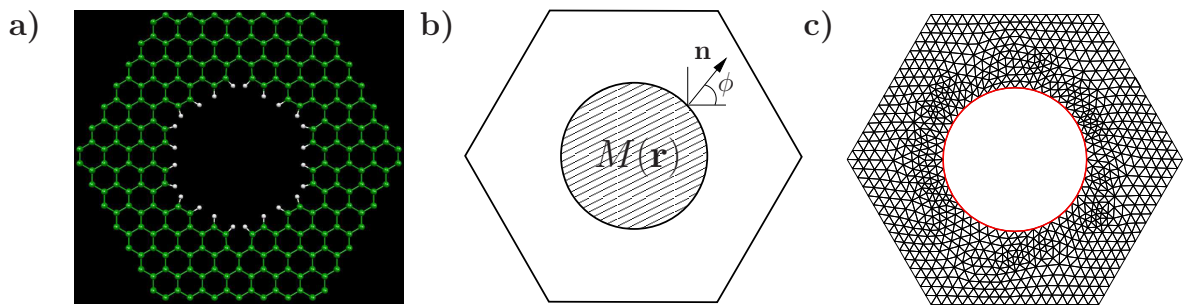}
\caption{Unit cell, continuum domain, and finite-element mesh. {\bf a)} Hexagonal unit cell of the $\{7,3\}$ graphene antidot lattice. {\bf b)} Corresponding continuum domain on which the Dirac equation is solved. The hole (hatched area) is modeled with a mass term $M(\mathbf{r})$ in the Dirac equation. The normal vector to the hole $\mathbf{n}$, forming the angle $\phi$ with the horizontal axis, is used to define appropriate boundary conditions along the edge of the hole (see text) {\bf c)} Corresponding finite-element mesh on which we solve the Dirac equation. The edge of the hole is shown with red. Periodic Bloch conditions are imposed on the outer boundary of the unit cell.}
\label{fig:diraceq}
\end{center}
\end{figure}

We first describe our finite-element solutions of the Dirac equation. The method is based on the band structure of bulk graphene close to the two Dirac points being linear and well described by the Dirac equation \cite{Neto2009}. Within this picture, the atomic honeycomb lattice structure of graphene is replaced by an effective continuum description. As an example, we show in Figs.\ \ref{fig:diraceq}a and \ref{fig:diraceq}b, respectively, a graphene antidot lattice unit cell and the corresponding continuum domain on which the Dirac equation is solved. The hole in the unit cell is mimicked with a mass term $M(\mathbf{r})$ in the Dirac equation at the location of the hole; see explanation following Eq.\ (\ref{eq:diracrealspace}). For large masses, the Dirac fermions are effectively excluded from the location of the hole and the mass term can be replaced by appropriate boundary conditions along the edge of the hole, indicated with red in Fig.\ \ref{fig:diraceq}c. In Fig.\ \ref{fig:diraceq}c we also show an example of the finite-element mesh on which the Dirac equation is discretized and solved. Periodic Bloch boundary conditions are imposed on the outer edges of the unit cell, making the problem equivalent to that of an infinitely large graphene antidot lattice.

Electronic states close to one of the two Dirac points of bulk graphene can be expressed in terms of envelope wave functions contained in the two-component spinor $|\Psi\rangle$ with one component corresponding to each of the two sublattices in the honeycomb structure of graphene \cite{Neto2009}. Spinors corresponding to states close to one of the Dirac points satisfy the Dirac equation
\begin{equation}
\hat{H}|\Psi\rangle=\left[\upsilon_F\,\hat{\mathbf{p}}\!\cdot\!\hat{\bm{\sigma}}+M(\hat{\mathbf{r}})\hat{\sigma}_z\right]|\psi\rangle=E|\Psi\rangle,
\label{eq:dirac}
\end{equation}
where $\upsilon_F\simeq 10^6$ ms$^{-1}$ is the Fermi velocity \cite{Novoselov2005}, $\hat{\mathbf{p}}=[\hat{p}_x,\hat{p}_y]$ is the momentum, $\hat{\bm{\sigma}}=[\hat{\sigma}_x,\hat{\sigma}_y]$ is the pseudo-spin corresponding to the two sublattices, and $M(\hat{\mathbf{r}})$ is the mass that couples to $\hat{\sigma}_z$ and is non-zero only inside the holes.  Spinors associated with the other Dirac point satisfy Eq.\ (\ref{eq:dirac}) with the replacement $\hat{\bm{\sigma}}\rightarrow\hat{\bm{\sigma}}^{*}=[\hat{\sigma}_x,-\hat{\sigma}_y]$. Within this description, states close to different Dirac points are assumed not to couple. The real-space representation of the spinor $|\Psi\rangle$ is $\Psi(\mathbf{r})\equiv\langle\mathbf{r}|\Psi\rangle=[\psi_1(\mathbf{r}),\psi_2(\mathbf{r})]^T$, where $\psi_1$ and $\psi_2$ are the envelope functions corresponding to each of the two sublattices. Equation (\ref{eq:dirac}) is correspondingly written
\begin{equation}
\left[
  \begin{array}{cc}
     M(\mathbf{r})& -i\hbar\upsilon_F(\partial_x-i\partial_y) \\
    -i\hbar\upsilon_F(\partial_x+i\partial_y) & -M(\mathbf{r}) \\
  \end{array}
\right]
\left[
  \begin{array}{c}
    \psi_1(\mathbf{r}) \\
    \psi_2(\mathbf{r}) \\
  \end{array}
\right]
=
E
\left[
  \begin{array}{c}
    \psi_1(\mathbf{r}) \\
    \psi_2(\mathbf{r}) \\
  \end{array}
\right].
\label{eq:diracrealspace}
\end{equation}
We now consider the situation, where Dirac fermions are excluded from the holes, by taking the limit $M(\mathbf{r})\rightarrow\infty$ inside the holes. In that limit, we can derive the appropriate boundary conditions for the spinor along the edges of a hole and solve the resulting problem outside the holes. The boundary conditions are derived by requiring that no particle current runs into a hole.  The particle current operator is $\hat{\mathbf{j}}\equiv\bm{\nabla}_{\bm{\hat{\mathbf{p}}}}\hat{H}=\upsilon_F\,\hat{\bm{\sigma}}$, and the local particle current density in the state $\Psi(\mathbf{r})$ is $\mathbf{j}(\mathbf{r})= \Psi^\dagger(\mathbf{r})\hat{\mathbf{j}}\Psi(\mathbf{r})$. Imposing $\mathbf{n}\cdot \mathbf{j}(\mathbf{r})=0$ along the edge of a hole with $\mathbf{n}$ being the outward-pointing normal vector to the hole, one can derive the condition $\psi_1(\mathbf{r})=ie^{-i\phi}\psi_2(\mathbf{r})$ along the boundary, where the angle $\phi$ is defined in Fig.\ \ref{fig:diraceq}b. This procedure was originally developed by Berry and Mondragon in studies of neutrino billiards \cite{Berry1987} and more recently employed by Tworzyd{\l}o and co-workers in the context of graphene \cite{Tworzydlo2006}. Along the outer boundaries of the unit cell we impose periodic Bloch boundary conditions, and we are thus left with a system of coupled differential equations on a finite-size domain with well defined boundary conditions. Problems of this type are well suited for commercially available finite-element solvers, and the
numerical implementation is relatively straightforward and fast using the standard finite-element package COMSOL Multiphysics \cite{Comsol}. The finite-element solver discretizes and solves the problem on an optimized mesh of the finite-size domain. The mesh shown in Fig. \ref{fig:diraceq}c was generated with COMSOL Multiphysics.

\subsection{Tight-binding (TB)}

We next describe our tight-binding scheme. The Dirac equation approach introduced above is a continuum description of the electronic properties, ignoring the detailed atomic structure of graphene and the edges of the holes, which may lead to scattering between the two Dirac points. It moreover assumes linear bands of bulk graphene. To capture effects of the atomic structure, including the influence of edge geometry, and in order to incorporate a realistic description of the band structure of bulk graphene, we need to go beyond the simple Dirac fermion picture. In our tight-binding scheme, the starting point is the Schr{\"o}dinger equation for a single electron in real-space representation
\begin{equation}
H^{\rm TB} \psi(\mathbf{r})=\left[-\frac{\hbar^2}{2m_e}\nabla^2 + V(\mathbf{r})\right]\psi(\mathbf{r}) = \epsilon \psi(\mathbf{r}),
\label{eq:TB}
\end{equation}
where $V$ is an effective potential and $m_e$ is the electron mass. The unknown eigenstate $|\psi\rangle$ is subsequently expanded in a set of localized ``atomic'' wave functions $|\vec{R},l\rangle$ as a superposition $|\psi\rangle=\sum C_{\vec{R },l}|\vec{R},l\rangle$ with expansion coefficients $C_{\vec{R },l}$. Here, each atomic state is labeled by the orbital symmetry ($l$=$s, p_x, p_z$...) and the position of the atom $\vec{R}$. This transforms the Schr{\"o}dinger equation
into a matrix equation reading
\begin{equation}
\sum_{\vec{R}',l'}\langle{\vec{R},l}| H^{\rm TB}|\vec{R}',l'\rangle C_{\vec{R}',l'}= \epsilon \sum_{\vec{R}',l'}\langle{\vec{R},l}|\vec{R}',l'\rangle C_{\vec{R}',l'}.
\label{eq:TB2}
\end{equation}
At this point, several approximations can be adopted in order to simplify the calculations. First, the atomic orbitals are usually taken to be orthogonal, i.e., $\langle{\vec{R},l}|\vec{R}',l'\rangle =\delta_{\vec{R},\vec{R}'}\delta_{l,l'}$. This means that the matrix problem becomes a simple rather than a generalized eigenvalue problem. Second, the matrix elements of $H^{\rm TB}$ are regarded as empirical parameters fitted, usually, to experimental data. In the simplest tight-binding description of planar carbon structures contained in the $(x,y)$-plane, just a single $p_z$ or $\pi$-orbital on each site is considered and only nearest-neighbor matrix elements are retained. This ``hopping integral'' is denoted as $-\beta$, with $\beta\approx 3.033$ eV~\cite{Saito1998}. Other values of the hopping integral can also be found in the literature. For example, the choice $\beta\approx 2.7$ eV provides low-energy band structures for bulk graphene consistently with density functional theory calculations \cite{Reich2002}. However, the Fermi velocity is determined by the relation $\upsilon_F=\sqrt{3}\beta a/2\hbar$ and by choosing $\beta\approx 3.033$ eV, we obtain $\upsilon_F=9.9 \times 10^5$ ms$^{-1}$ in good agreement with experiments \cite{Novoselov2005}.

The reason for considering only $\pi$-orbitals is that $\pi$-orbitals with odd $z$-parity decouple from the $\sigma$-orbitals spanned by $s, p_x$, and $p_y$ states that all have even $z$-parity. Moreover, the bands in the vicinity of the band gap are all produced by the loosely bound $\pi$-orbitals. Hence, for all structures considered in the present work, we need only include $\pi$-orbitals explicitly. Also, even though realistic structures will contain hydrogen terminated edges, the hydrogen atoms couple only to the $\sigma$-orbitals and are therefore irrelevant for $\pi$-states. In a more sophisticated model, bare or hydrogen terminated edges lead to a small modification of the $\pi$-electron hopping integrals near an edge due to relaxation of the geometry. This modification is ignored as it simply leads to a small additional opening of the band gap \cite{Pedersen2008}.

\subsection{Density functional theory (DFT)}

Finally, we discuss our DFT calculations. This method provides the most detailed description of graphene antidot lattices, and we expect it to yield the most accurate results. The accuracy comes at the cost of the method being numerically demanding and the required computational resources exceed those typically available on a standard PC. Density functional theory is a widely used standard for electronic structure calculations and we shall here only briefly outline the underlying theory \cite{Martin2004}.

The method takes as starting point the full interacting many-body system involving all electrons and atom nuclei making up the graphene antidot lattice. Diagonalizing the corresponding many-body Hamiltonian is a tremendous task, but the problem can be brought to a somewhat simpler form using the Born-Oppenheimer approximation in which the positions of the nuclei are fixed. We are then considering a system of interacting electrons moving in an external potential created by the nuclei at fixed positions. This is still a difficult many-body problem, but further advances can be made following Hohenberg and Kohn who showed that the ground state energy is uniquely determined by the ground state electron density \cite{Hohenberg1964}. Kohn and Sham (KS) later realized that this density can be obtained from a single-particle picture of non-interacting electrons. The corresponding Hamiltonian for the single-particle KS orbitals $\psi_i$ is expressed by the KS equations as~\cite{Kohn1965}
\begin{equation}
H^{\rm KS}\psi_i(\mathbf{r})=\left[-\frac{1}{2}\nabla^2 + V_{\rm eff}(\mathbf{r})\right]\psi_{i}(\mathbf{r}) = \epsilon_{i} \psi_{i}(\mathbf{r}),
\label{eq:KS}
\end{equation}
where the effective potential
\begin{equation}
V_{\rm eff}(\mathbf{r}) = \int\! d\mathbf{r}^{'}\frac{\rho(\mathbf{r}^{'})}{|\mathbf{r} -\mathbf{r}^{'}|}  + V_{a}(\mathbf{r},\{\mathbf{R}_{i_a}\}) + \frac{\delta E_{xc}[\rho(\mathbf{r})]}{\delta \rho(\mathbf{r})}
\end{equation}
depends explicitly on the density $\rho(\mathbf{r}) = \sum_{i_{o}}|\psi_i(\mathbf{r})|^2$ with the sum running over occupied KS orbitals. Here, $V_{a}(\mathbf{r},\{\mathbf{R}_{i_a}\})$ is the external potential due to the atoms at positions $\mathbf{R}_{i_a}$. The so-called exchange-correlation term $E_{xc}(\mathbf{r})$ accounts for all many-body effects and is not known exactly, but must be appropriately approximated. Finally, the ground state energy of the interacting problem is
\begin{eqnarray}
E[\rho(\mathbf{r})] = & T[\rho(\mathbf{r})] + \int \! d\mathbf{r} \, \rho(\mathbf{r})  V_{a}(\mathbf{r},\{\mathbf{R}_{i_a}\}) \nonumber\\
 &+ \frac{1}{2}\int\!\!\int d\mathbf{r}d\mathbf{r}^{'}\,\frac{\rho(\mathbf{r})\rho(\mathbf{r}^{'})}{|\mathbf{r} -\mathbf{r}^{'}|} + E_{xc}[\rho(\mathbf{r})],
\end{eqnarray}
where $T$ is the kinetic energy corresponding to the density $\rho(\mathbf{r})$.

We are now left with the problem of determining the density $\rho(\mathbf{r})$. The density is a function of only three coordinates, unlike the $N$-particle wavefunction of $3N$ coordinates. The set of KS equations is solved self-consistently: starting from an initial density, the effective potential is computed together with the KS orbitals and the corresponding density, and this procedure is repeated until convergence has been reached. The band structure can then by calculated corresponding to the chosen coordinates of the nuclei $\mathbf{R}_{i_a}$. The total energy of the system can further be minimized with respect to the coordinates of the nuclei. This is referred to as geometry relaxation. The method can easily be extended to include spin as well as different species of nuclei. In this work, we use spin-polarized DFT as implemented in the \siesta\ code \cite{Soler2002}. The structures are relaxed using computationally cheaper DFT based tight-binding methods \cite{Porezag1995}. Performing electronic structure calculations using DFT on geometries relaxed in this way is known to provide accurate results \cite{Fuerst2009}.
As commonly done, the core electrons are replaced by pseudo-potentials and the remaining valence-electrons are described with localized atomic orbitals.  For the exchange-correlation potential we employ the widely used Perdew-Burke-Ernzerhof parametrization of the generalized gradient approximation \cite{Perdew1996}. We mainly use a so-called double-$\zeta$ polarized (DZP) basis set size, consisting of 13 functions per carbon atom. Contrary to the DE and TB methods, the antidot edges are hydrogen-passivated in the DFT calculations. The effects of passivation are discussed in Section \ref{sec:passivation}. Further details of our DFT calculations can be found in Ref.\ \cite{Fuerst2009}.

\begin{figure}
\begin{center}
\includegraphics[width=.9\linewidth]{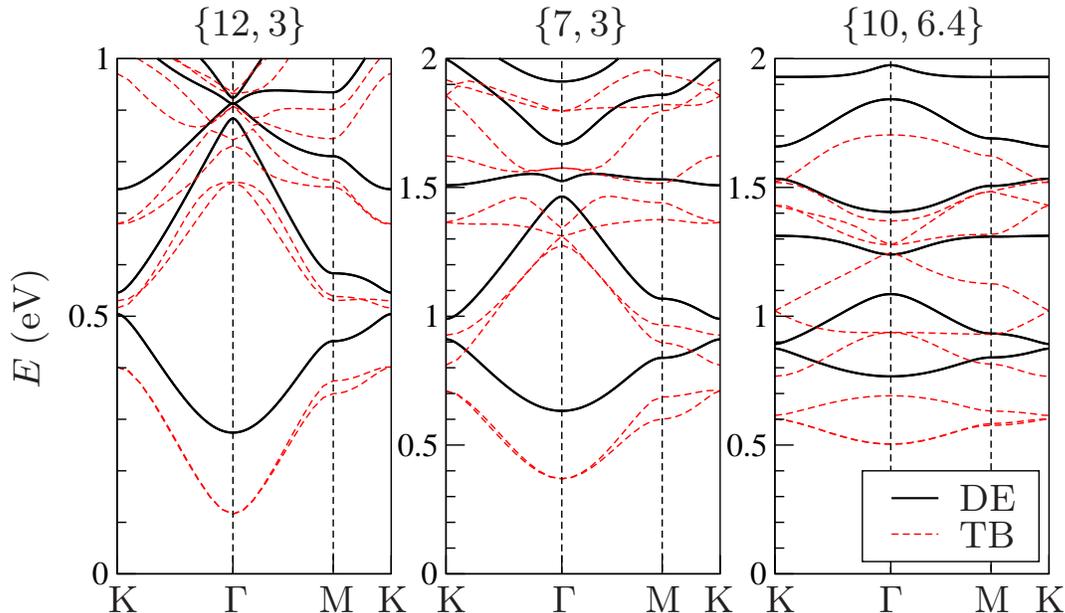}
\caption{Band structures of three representative graphene antidot lattices. Full lines indicate results obtained by solving the Dirac equation (DE), while tight-binding results (TB) are shown with red dashed lines. Within these computational approaches we have exact particle-hole symmetry, and consequently only positive energies are shown. Note the different energy scale on the leftmost figure.}
\label{fig:DEvsTB}
\end{center}
\end{figure}

\section{Band structures}
\label{sec:bandstruc}

We now present and compare our results for the electronic band structure obtained using the three methods described in the previous section. This provides valuable insight into the physics dominating the electronic properties of graphene antidot lattices as well as an indication of the range of validity of the less computationally expensive methods. Both the finite-element solutions of the Dirac equation (DE) and our tight-binding calculations (TB) were carried out on a standard PC, and a single band structure calculation could typically be performed in a few minutes for the relatively small-scale graphene antidot lattices considered in the following. The density functional theory calculations (DFT) were carried out on 8 AMD Opteron CPUs in parallel and typically lasted around 48 hours. Unlike the TB and the DFT methods, the computational time of our DE scheme does not increase with the size of the unit cell, determined by $L$, but only depends on the ratio $R/L$. For large unit cells, the DE scheme will therefore outperform both the TB and the DFT methods in terms of computational time.

In Fig.~\ref{fig:DEvsTB} we show band structure results for three representative graphene antidot lattices using DE and TB. Both methods predict band gaps of a few hundred meVs for these relatively small dimensions of graphene antidot lattices. For low energies, DE predicts well the qualitative features of the bands obtained using TB, but the deviations become pronounced at higher energies. This is not surprising as the Dirac equation is only a valid description at low energies, where the band structure of bulk graphene is linear. Roughly, this means energies below $0.1\beta\simeq 0.3$ eV. Additionally, the increased kinetic energy due to the confinement of the particles renders the DE results less accurate for large antidot radii relative to the dimensions of the unit cell. This is apparent in the figure, where the bands at higher energies become increasingly inaccurate as the antidot radius is increased. However, even for the $\{L,R\}=\{10, 6.4\}$ structure, there is a qualitative agreement between the shapes of the bands at low energies found using the two methods.

\begin{table}
\begin{center}\begin{tabular}{|l|c|c|c|c|c|c|}
\hline
\hline
&  \multicolumn{2}{|c|}{$\{12,3\}$}   &   \multicolumn{2}{|c|}{$\{7,3\}$}   &   \multicolumn{2}{|c|}{$\{10,6.4\}$}\\
\hline
& eV & $\Delta_{\{12,3\}}$ & eV & $\Delta_{\{12,3\}}$ & eV & $\Delta_{\{12,3\}}$ \\
\hline
DE & 0.54 (0.29) & 1 & 1.27 (0.82) & 2.35 (2.83) & 1.53 (1.22) & 2.83 (4.21)\\
\hline
TB  &  0.23 & 1 & 0.74 & 3.22 & 1.01 & 4.39\\
\hline
DFT  &  0.19 & 1 & 0.61 & 3.21 & 0.82 & 4.32
\\
\hline
\hline
\end{tabular}
\end{center}
\caption{Band gaps of three representative graphene antidot lattices. We show results obtained by solving the Dirac equation (DE), via tight-binding calculations (TB), and using density functional theory (DFT). Values in parentheses are obtained using DE and corrected for the low-radius behavior (see text). The band gaps are given in eV as well as in dimensionless values relative to the size of the band gap for the $\{L,R\}=\{12,3\}$ structure.}
\label{bandgaps}
\end{table}

While the shapes of the bands at low energies are approximately the same for the DE and TB approaches, the sizes of the band gaps vary significantly. For the $\{L,R\}=\{12,3\}$ structure, the band gap predicted by DE is more than twice
as large as that obtained using TB. These differences may be traced back to two of the underlying assumptions of DE: linear bands of bulk graphene and absence of scattering between the two Dirac points. In order to illuminate the discrepancy we consider two limiting cases. We first consider the limit of large unit cells, i.e., large values of $L$. By investigating a large sample of different graphene antidot lattice using TB, some of us have demonstrated a simple scaling-law between the hole size and the band gap $E_g$, showing that $E_g \propto \sqrt{N_{\rm hole}}/N_{\rm cell}$ for small values of the ratio $R/L$ \cite{Pedersen2008}. Here, $N_{\rm  hole}\propto R^2$ is the number of carbon atoms that have been removed from the intact unit cell in order to create the hole, and $N_{\rm cell}\propto L^2$ is the total number of carbon atoms in the intact unit cell (before the hole is made). We then find  $E_g \propto (R/L)/L$, showing that for a fixed value of the geometric ratio $R/L$, the band gap $E_g$ falls off as $1/L$ with increasing $L$. For sufficiently large unit cells we thus expect the band gap to be well within the energy window for which the electronic bands of bulk graphene in fact are linear, and the band gaps obtained using DE should thus agree better with TB. The limit of small holes, i.e., $R$ going to 0, is another important check point of our methods.  In the DE approach, the boundary condition along the edge of a hole enforces a phase shift between the two spinor components, given by the angle $\phi$ indicated in Fig.\ \ref{fig:diraceq}b. With $R$ going to 0, the phase shift must occur at a single point in space, resulting in a completely undetermined phase relationship at this point. Consequently, there is no adiabatic transition from a graphene antidot lattice to bulk graphene in the limit of vanishing hole sizes. Indeed, for small values of $R$, we find a non-vanishing band gap using DE, and extrapolating the results to $R=0$ we find a band gap of the approximate size $1.02\beta/L$. If we correct for this by simply subtracting this value from the band gaps calculated using DE, we find better agreement with the results obtained using TB, as shown in Table \ref{bandgaps}. For large values of $L$, the correction tends to zero, as we would expect.

\begin{figure}
\begin{center}
\includegraphics[width=.9\linewidth]{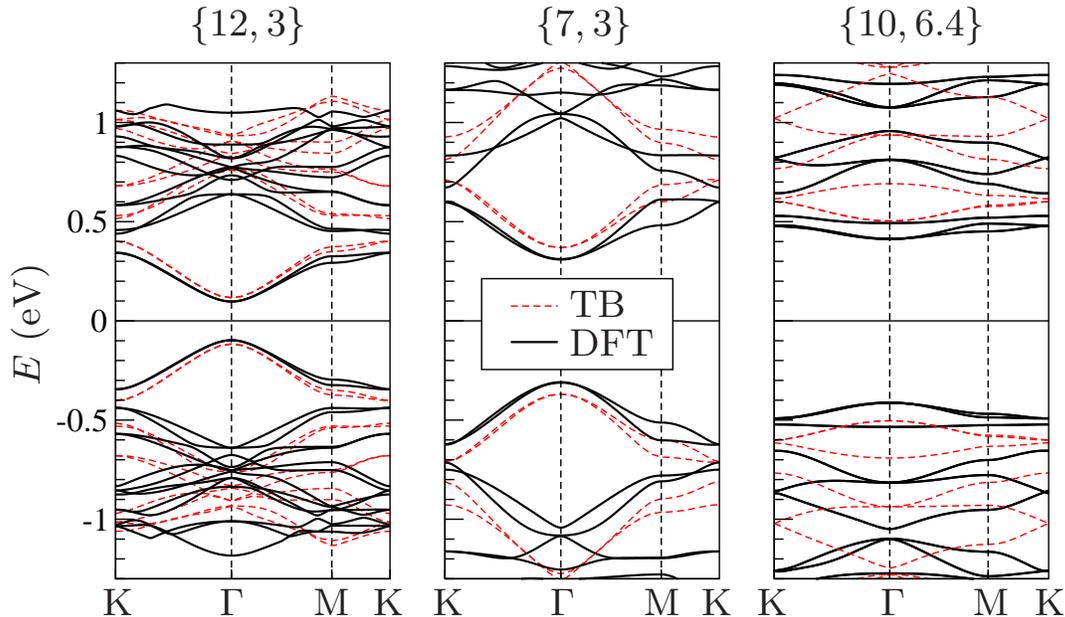}
\caption{Band structures of three representative graphene antidot lattices. Full lines indicate results obtained using density functional theory (DFT), while tight-binding results (TB) are shown with red dashed lines. Within the DFT scheme, particle-hole symmetry is not assumed, and we thus show results for energies both above and below the Fermi energy at zero.}
\label{fig:dftvstb}
\end{center}
\end{figure}

In Table \ref{bandgaps} we also show results for the band gaps using density functional theory. The band gaps calculated using DFT are within 30\% of the corresponding TB results, with DFT consistently reporting lower band gaps than TB. This follows the general tendency that energy gaps are underestimated in DFT \cite{Onida2002}. For the three structures shown here, the band gaps increase with increasing relative hole size. This trend is captured well by all three methods. The band structures calculated with DFT are shown in Fig.\ \ref{fig:dftvstb}, together with results obtained using TB for comparison. Generally, there is a reasonable qualitative agreement between the two methods in terms of the shapes of the bands, in particular, at energies close to the Fermi level. At larger energies, the qualitative features start to deviate significantly. Unlike the TB calculations, the DFT approach does not imply particle-hole symmetry, and the corresponding band structures are not symmetric around the Fermi level at zero. This difference is clearly seen in the figure.

\begin{figure}
\begin{center}
\includegraphics[width=.9\linewidth]{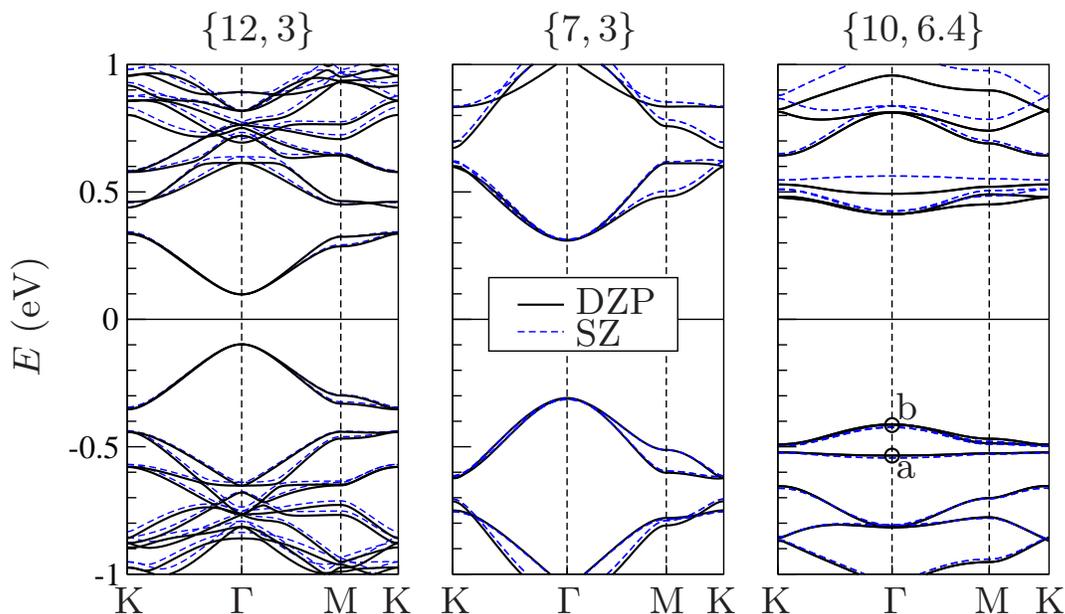}
\caption{Band structures of three representative graphene antidot lattices calculated with DFT. Full lines indicate results obtained using the DZP basis set, while dashed lines correspond to the smaller SZ basis set (see text). The real-space representations of the states corresponding to the points $a$ and $b$ are shown in Fig.\ \ref{fig:realspacerep}.}
\label{fig:DFTcomparison}
\end{center}
\end{figure}

In contrast to the DE and TB calculations, our DFT  scheme also includes the spin degree of freedom and is thereby able to predict the magnetic properties of the graphene antidot lattice. Within the TB description, graphene is considered a bipartite lattice structure with two sublattices, $A$ and $B$, with non-zero hopping elements between different sublattice sites only. In that case, the total magnetic moment per unit cell $M$, can be determined from Lieb's theorem \cite{Lieb1989}, stating that $M=N_{A}-N_{B}$ with $N_{A(B)}$ being the number of sites of sublattice $A(B)$ in the unit cell. By inspection of the structures in Fig.\ \ref{fig:unitcells}, we see that they have zero sublattice imbalance, i.e., $N_A=N_B$, and we thus expect a zero total magnetic moment according to Lieb's theorem. Although our DFT calculations are not based on a description of graphene in terms of two sublattices with only nearest-neighbor coupling, we still find a zero total magnetic moment. Additionally, we find that no local magnetic moments are formed in any of the investigated structures. Lieb's theorem does not concern the formation of local magnetic moments, but the absence in the present cases can be understood from the circular shapes of the holes, which inhibit the formation of longer zig-zag shaped parts of the edge. This is similar to results obtained for graphene flakes, where relatively large zig-zag parts are needed for local magnetic moments to form \cite{Jiang2007}. We thus conclude that the bands are all spin degenerate in the cases we have investigated.

\begin{figure}
\begin{center}
\includegraphics[width=0.75\linewidth]{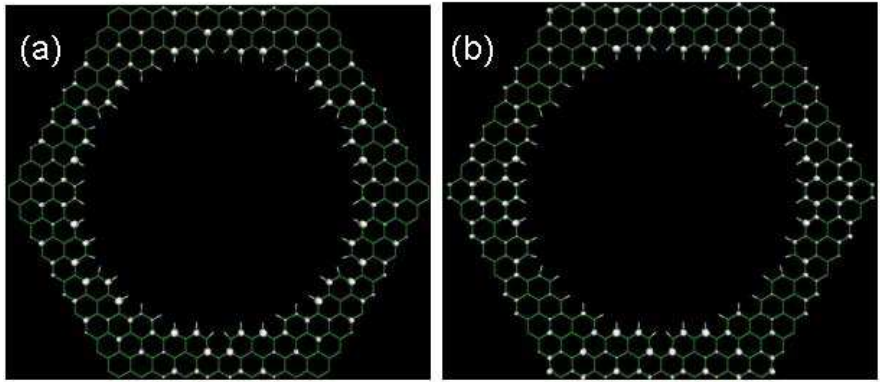}
\end{center}
\caption{Real-space representation of electronic states. The left panel corresponds to the point on the flat band in Fig.\ \ref{fig:DFTcomparison} indicated by $a$. For comparison, the right panel shows the state corresponding to the point $b$ in Fig.\ \ref{fig:DFTcomparison}. We show the absolute square of the wavefunctions.}
\label{fig:realspacerep}
\end{figure}

Within our DFT scheme, the computational time can be reduced by using a smaller basis set. We thus compare results obtained with the double-$\zeta$ polarized (DZP) basis set involving 13 basis functions per carbon atom, used thus far, and results obtained using a single-$\zeta$ (SZ) basis with only 4 basis functions per carbon atom. Results for the band structures obtained using the two different basis sets are shown in Fig.\ \ref{fig:DFTcomparison}. The band structures obtained using the smaller SZ basis agree well with those obtained using the DZP basis, and the computational time is significantly reduced. An interesting difference between the band structures obtained using DFT compared to DE and TB, is the very low dispersion of the band roughly 0.5 eV below the Fermi level for the $\{L,R\}=\{10,6.4\}$ structure. The absolute square of the wavefunction for one of the spin degenerate states at the $\Gamma$-point, denoted by $a$ in Fig.\ \ref{fig:DFTcomparison}, is shown in Fig. \ref{fig:realspacerep}. For comparison, we also show the state denoted by $b$ in Fig.\ \ref{fig:DFTcomparison}. The state on the flat band is  strongly localized on the zig-zag parts of the edge. The lower dispersion compared to TB is possibly due to the gradually increasing total electronic potential within DFT, when approaching the edge of a hole. The increased on-site energy of the edge atoms within DFT may thus cause stronger localization.

\section{Passivation}
\label{sec:passivation}

\begin{figure}
\begin{center}
\includegraphics[width=0.75\textwidth]{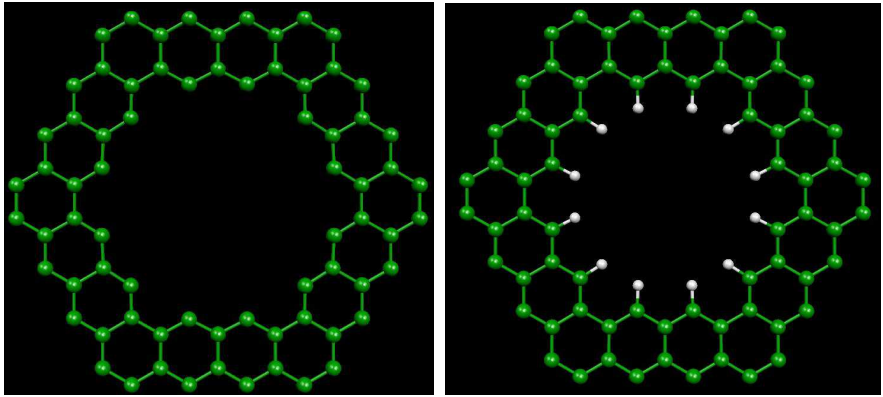}
\end{center}
\caption{The unit cell of the $\{L,R\}=\{4,2\}$ structure. The structure is shown with (right) and without (left) complete hydrogen passivation of the carbon atoms along the edge of the hole.}
\label{fig:passivation}
\end{figure}

Finally, we discuss the influence of edge passivation of the holes with hydrogen. In order to address this question we employ our DFT scheme. Details of the edges are not considered within our finite-element solutions of the Dirac equation (DE), and within a tight-binding description (TB), passivation is typically included  simply as a shift of the hopping integral between carbon atoms along the edges due to the relaxed carbon-carbon bond length \cite{Son2006}. This correction leads to slightly increased energy gaps but has been ignored in the TB calculations in the present work.
In contrast, DFT carefully treats the presence of hydrogen along the edge of a hole, and, importantly, the method includes the spin degrees of freedom, which turns out to be crucial in determining the influence of passivation on the electronic properties. We consider as an illustrative example the structure $\{L,R\}=\{4,2\}$ depicted in Fig.\ \ref{fig:passivation}, shown with and without hydrogen passivation. We note that the hole geometry in this case is hexagonal, leading again to a vanishing magnetic moment without passivation.

\begin{figure}
\begin{center}
\includegraphics[width=0.65 \columnwidth]{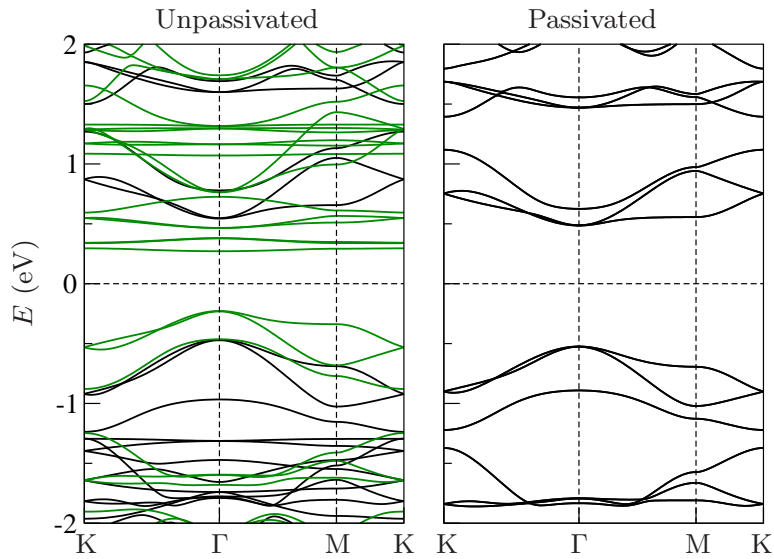}
\end{center}
\caption{Band structure of the $\{L,R\}=\{4,2\}$ graphene antidot lattice calculated with DFT. The left (right) panel shows the band structure without (with full) hydrogen passivation, corresponding to the unit cells in Fig.\ \ref{fig:passivation}. The systems are fully relaxed and the spin degree of freedom is included. Majority (minority) spin is shown with black (green). The Fermi level is at $E=0$.}
\label{fig:passivationbands}
\end{figure}

The resulting band structures, shown in Fig.\ \ref{fig:passivationbands}, with and without passivation are very different. With full hydrogen passivation, several bands are spin degenerate. This degeneracy is lifted without passivation, and low dispersion bands stemming from dangling bonds are clearly present. The dispersions of these bands arise due to coupling between neighboring edge atoms. Each dangling bond introduces a calculated spin of one Bohr magneton, 1.00$\mu_B$, giving a total magnetization of 12.00$\mu_B$ per cell, causing the lifting of the spin degeneracy. This magnetization involves only the $sp^2$-orbitals and is strongly localized at the sites of the dangling bonds. We find that structural relaxation has no qualitative impact on the results in these two cases.

\begin{figure}
\begin{center}
\includegraphics[width=0.78 \columnwidth]{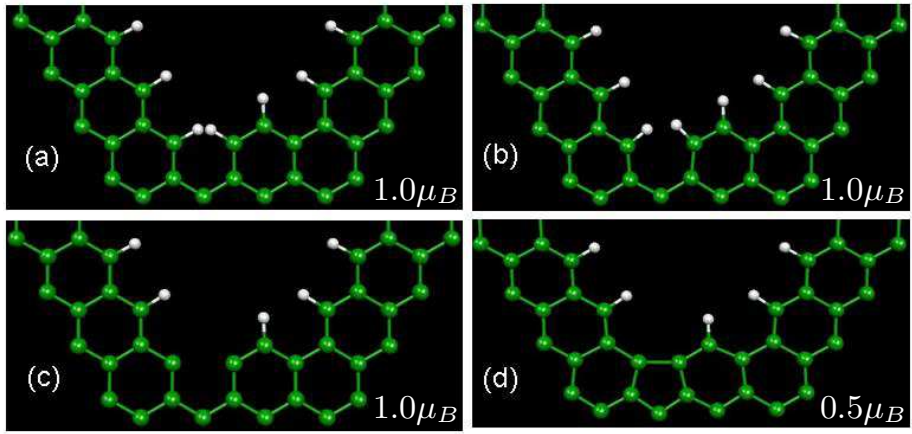}
\end{center}
\caption{Single carbon vacancy at the edge of the hole in the $\{L,R\}=\{4,2\}$ structure. In the left (right) panels
the structures have not been (have been) relaxed. In the upper (lower) panels, the carbon atoms next to the vacancy have (have not) been passivated with hydrogen. The calculated magnetic moment is indicated in each panel.}
\label{fig:defectgeom}
\end{figure}

\begin{figure}
\begin{center}
\includegraphics[width=0.65 \columnwidth, trim = 0 0 0 0, clip]{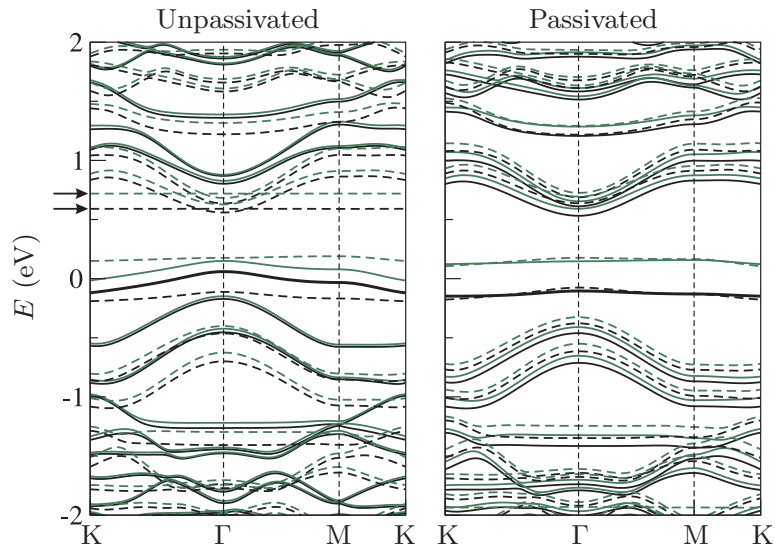}
\end{center}
\caption{Band structures of the $\{L,R\}=\{4,2\}$ graphene antidot lattice with a single carbon vacancy in the unit cell. Dashed lines indicate band structures for the unrelaxed geometry shown in Fig.\ \ref{fig:defectgeom}, panels (a) and (c), while full lines are the corresponding results for the relaxed structures, panels (b) and (d). The
unfilled bands of the dangling bonds are indicated in the left panel by horizontal arrows. The corresponding filled bands at lower energies are not shown in the plot. Majority (minority) spin is shown with black (green). The Fermi level is at $E=0$.}
\label{fig:pentagonbands}
\end{figure}

Next, we investigate the effects of a single carbon vacancy at the edge. This introduces a sublattice imbalance of $|N_A-N_B|=1$, resulting in an expected non-zero total magnetic moment according to Lieb's theorem. Elaborating on Lieb's work, Inui, Trugman and Abraham have shown that such a sublattice imbalance is accompanied by at least $|N_A-N_B|$ midgap states with zero energy for a perfect bipartite lattice \cite{Inui1994}. A recent discussion of similar statements can be found in Ref.\ \cite{Wang2009}. In Fig.\ \ref{fig:defectgeom} we show the geometries of a single carbon vacancy at the edge, both with and without hydrogen passivation, as well as with and without having relaxed the geometries. The corresponding band structures are shown in Fig.\ \ref{fig:pentagonbands}. Generally, we find two low dispersion bands close to the Fermi level. These are the midgap states with an induced spin splitting. The spin degeneracy is lifted for all bands due to the non-zero total magnetic moment. The main finding in the case without passivation of the atoms close to the vacancy are the two flat bands stemming from the dangling bonds, indicated with arrows in Fig.\ \ref{fig:pentagonbands}. The dangling bonds are found to overlap in case of which it is energetically most favorable for the dangling bonds to have zero total spin as our calculations show. We find a magnetization of $1.00\mu_B$ per unit cell for both systems in the unrelaxed case. This magnetization is entirely due to the sublattice imbalance, and is, contrary to the case of dangling bonds, largely non-local, residing mainly on the $\pi$-orbitals.

Whereas relaxation has minor effects when passivation is included, the opposite is true for carbon vacancies without passivation. In that case, the two unpassivated carbon atoms at the edge approach each other, forming a pentagon as seen in Fig. \ref{fig:defectgeom}d. A similar phenomenon has been observed theoretically for single carbon vacancies in bulk graphene, where the spin of such vacancies can usually be understood as an unsaturated dangling bond on the neighboring carbon atom, not forming the pentagon \cite{Ma2004,Lehtinen2004,Yazyev2007}. This results in a calculated magnetic moment of around $1\mu_B$.  In our case, however, there are only two neighboring atoms. In fact, in both cases the magnetic moment is better understood using Lieb's theorem, as discussed by Palacios, Fern\'{a}ndez-Rossier, and Brey, in the case of carbon vacancies in bulk graphene \cite{Palacios2008}. In the pentagon geometry, two sites belonging to the same sublattice bond stronger to each other, which is reflected in the smaller bond length of 1.67~\AA\ compared to 2.46~\AA\ in the case without relaxation. Consequently, the dangling bonds are then saturated and the corresponding flat bands are not present. Additionally, the bipartite lattice symmetry is broken, causing a reduction in the magnetic moment from $1.00\mu_B$ to $0.50\mu_B$. Consequently, the spin splitting of the bands is reduced. The midgap states are still observed, but in this case with more dispersion. The features of the bipartite lattice are thus maintained in a moderated version, when pentagons are formed due to a carbon vacancy along the edge of the hole. We stress that while the magnetization arising from Lieb's theorem is predictable, the magnetization due to dangling bonds is merely a result of energy optimization and therefore harder to predict.

\section{Conclusions}
\label{sec:conclusions}

We have carried out a numerical study of the band structures of graphene antidot lattices, using three different computational approaches of varying levels of complexity and accuracy. Finite-element solutions of the Dirac equation (DE) provide a simple and fast scheme, capturing essential qualitative features of the band structures and band gaps. For more reliable predictions of the band structures, we employed a $\pi$-orbital tight-binding scheme (TB) as well as computationally heavy density functional theory calculations (DFT). The three methods all predict an opening of a band gap on the order of a few hundred meVs for the nano-scale structured graphene antidot lattices studied in this work. Qualitative similarities were found for the band structures calculated with the three different methods. Finally, we discussed the effects of hydrogen passivation along the edges of the holes. Passivation was found to have a significant influence on the band structures, and the presence of carbon vacancies along the hole edges were shown to induce midgap bands.

\section{Acknowledgments}

We thank B Trauzettel, J Li, M Vanevi\'{c}, and V M Stojanovi\'{c} for insightful discussions. The work by CF was supported by the Villum Kann Rasmussen Foundation. Financial support from Danish Research Council FTP
grant `Nanoengineered graphene devices' is gratefully acknowledged. Computational resources were provided by
the Danish Center for Scientific Computations (DCSC). APJ is grateful to the FiDiPro program of the Finnish
Academy.

\newpage
\section*{References}

\bibliographystyle{iop}

\end{document}